\documentclass[%
 reprint,
superscriptaddress,
 amsmath,amssymb,
 aps,
prb,
floatfix,
longbibliography
]{revtex4-2}

\usepackage{silence}%https://tex.stackexchange.com/questions/180762/revtex4-1-warning-repair-the-float-package
\usepackage{graphicx}% Include figure files
\usepackage{dcolumn}% Align table columns on decimal point
\usepackage{bm}% bold math
\usepackage{bbm} % \mathbbm{1}
\usepackage{mathtools}
\usepackage{leftindex} % \leftindex https://tex.stackexchange.com/questions/11542/left-and-right-subscript-superscript
\usepackage[colorlinks=true,allcolors=blue]{hyperref}

\graphicspath{{./Figures01/}} 

\usepackage{float} % Paquete que permite la orden [H] = Pon esta figura aqu\'{i} bajo mi responsabilidad por muy feo que quede
\usepackage{fix-cm} % Caracteres de cualquier tamaño!!!
\usepackage[dvipsnames]{xcolor}
\usepackage{xcolor}
\usepackage[normalem]{ulem}
\usepackage[english]{babel}
\usepackage{graphicx}% Include figure files
\usepackage{dcolumn}% Align table columns on decimal point
\usepackage{verbatim}
\usepackage{mathrsfs}
\usepackage{cancel}
\usepackage{epstopdf}
\usepackage{color}
\usepackage{tensor}
\usepackage{relsize} % Escalar caracteres
\usepackage{ytableau}
\usepackage{slashed} % For Feynamnn slash notation in Dirac operators \slashed{p}
\usepackage{boldline} % For using V{3} and \hlineB{4}. https://tex.stackexchange.com/questions/425803/how-to-place-thick-vertical-line-in-table
\usepackage{csquotes} % Paquete para usar bien las comillas en Latex

\usepackage{pifont}% http://ctan.org/pkg/pifont
\usepackage[dvipsnames]{xcolor} 
\definecolor{Verde}{RGB}{0, 150, 0} %RGB de 0 a 255

\DeclareFontFamily{U}{mathc}{}
\DeclareFontShape{U}{mathc}{m}{it}%
{<->s*[1.03] mathc10}{}
\DeclareMathAlphabet{\mathscr}{U}{mathc}{m}{it}

\newcommand{\dd}{\mathrm{d}}
\newcommand{\ii}{\mathrm{i}}

\newcommand{\qq}{\mathrm{q}}
\newcommand{\kk}{\mathrm{k}}

\newcommand{\surf}{\parallel}
\newcommand{\dlim}{\displaystyle\lim}

\DeclareMathOperator{\ReOp}{Re}
\newcommand{\Real}[1]{\ReOp\!\left[#1\right]}
\DeclareMathOperator{\ImOp}{Im}
\newcommand{\Imag}[1]{\ImOp\!\left[#1\right]}
\newcommand{\Eq}[1]{Eq.~\eqref{#1}}

\newcommand{\Comment}[1]{\textit{\enquote{#1}}}

\begin{document}
%\linenumbers

\thispagestyle{empty}

\title{Reply to: Comment on ``Electric conductivity of graphene: Kubo model versus a nonlocal quantum field theory model" }

\author{Pablo Rodriguez-Lopez}
\email{pablo.ropez@urjc.es}
\affiliation{{\'A}rea de Electromagnetismo and Grupo Interdisciplinar de Sistemas Complejos (GISC), Universidad Rey Juan Carlos, 28933, M{\'o}stoles, Madrid, Spain}
\affiliation{Laboratoire Charles Coulomb (L2C), UMR 5221 CNRS-University of Montpellier, F-34095 Montpellier, France}

\author{Jian-Sheng Wang}
\affiliation{Department of Physics, National University of Singapore, Singapore 117551, Republic of Singapore}

\author{Mauro Antezza}
\email{mauro.antezza@umontpellier.fr}
\affiliation{Laboratoire Charles Coulomb (L2C), UMR 5221 CNRS-University of Montpellier, F-34095 Montpellier, France}
\affiliation{Institut Universitaire de France, Ministère de  l’Enseignement Supérieur et de la Recherche, 1 rue Descartes, F-75231, Paris, France}

\begin{abstract}
In the Comment by Bordag et al. \cite{CommentPRB}[Phys. Rev. B \textbf{113}, 207401 (2026)], concerns are raised regarding the validity of the results presented in \cite{PabloMauroComparisonKuboQFT2024} [Phys. Rev. B \textbf{111}, 115428 (2025)], where the theoretical descriptions of the electric conductivity of graphene obtained from the Kubo formula and from quantum field theory via the polarization tensor are compared. In this Reply, we show that these concerns arise from misinterpretations of our work, in which the results are either inaccurately represented or applied outside the domain of validity of the model. We address the comments concerning the derivation of the Luttinger formula for the electric conductivity from the Kubo formula and clarify why the results of our work cannot be arbitrarily extended to make claims on the gauge invariance. We further demonstrate that our findings are fully consistent with the established and widely accepted literature cited in the Comment. We confirm that the model for electric conductivity discussed in our work correctly predicts a vanishing electric current in the absence of an external electric field, as physically required, and in contrast with the model advocated by the Authors of the Comment. We also show that the electric permittivity does not exhibit a double pole in $\omega$, contrary to the claim made in the Comment. Finally, we emphasize that the inclusion of losses is a standard and well-established approach in the study of transport properties of materials, including graphene, and we take the opportunity to correct a few minor typographical errors in our work. We show and maintain that all results derived in our work are fully valid and correct.
\end{abstract}

\maketitle

%\section{Introduction}
The comment of Bordag et al.\cite{CommentPRB} (for the sake of brevity, hereafter referred to as ``the Comment'') lists concerns about the validity of the results of \cite{PabloMauroComparisonKuboQFT2024}.

%\section{Response to the criticisms}
We reply here to all the remarks one by one.
%\subsection{Correction of the relation between electric current and electromagnetic field}
The Comment claims that, in \cite{PabloMauroComparisonKuboQFT2024} the relation 
\begin{eqnarray}\label{Def_Pi}
J_{\mu} = - \Pi_{\mu\nu}A^{\nu}.
\end{eqnarray}
is considered as unsatisfactory, and hence it is modified by replacing $\Pi_{\mu\nu}$ with the modified expression $\tilde{\Pi}_{\mu\nu}(\omega,\bm{q}) = \Pi_{\mu\nu}(\omega,\bm{q}) - \dlim_{\omega\to0}\Pi_{\mu\nu}(\omega,\bm{q})$.

This is not the case, \Eq{Def_Pi} is correct. This is explicitly written in the very first section of \cite{PabloMauroComparisonKuboQFT2024} to avoid any misunderstanding. Moreover, we also explicitly use \Eq{Def_Pi} in our published works, like \cite{PabloMauro2025}, where the full electromagnetic response is needed. In \cite{PabloMauroComparisonKuboQFT2024} $\Pi_{\mu\nu}$ is considered as the transport coefficient that relates the induced electric 4-current with the full electromagnetic field, given by the 4-vector $A^{\nu}$. This point is uncontroversial.

What is stated in the first section of \cite{PabloMauroComparisonKuboQFT2024} is that, if one is interested in the purely electric conductivity, defined as the transport coefficient that relates the induced electric current $J_{\mu}$ to the electric field $E^{\nu}$ \emph{only} as
\begin{eqnarray}\label{OhmLaw}
J_{\mu} = \sigma_{\mu\nu}E^{\nu},
\end{eqnarray}
it is not correct to simply write it as 
\begin{eqnarray}\label{sigmaNR}
\sigma_{\mu\nu}^{\rm{NR}}(\omega,\bm{q}) = \dfrac{\Pi_{\mu\nu}(\omega,\bm{q})}{\ii\omega},
\end{eqnarray}
as done by the Authors of the Comment. This choice leads to unphysical results, such as the presence of electric currents described by the electric conductivity, but in the absence of an electric field. In \cite{PabloMauroComparisonKuboQFT2024} it has been rigorously shown that the correct expression for the electric conductivity is the well-known Luttinger formula derived from Kubo formalism
\begin{eqnarray}\label{sigmaK}
\sigma_{\mu\nu}^{\rm{K}}(\omega,\bm{q}) = \dfrac{\Pi_{\mu\nu}(\omega,\bm{q}) - \dlim_{\omega\to0}\Pi_{\mu\nu}(\omega,\bm{q})}{\ii\omega}.
\end{eqnarray}
\Eq{sigmaK} is hence mathematically re-derived in detail on the basis of causality; it predicts no electric currents in the absence of an electric field.

%\subsection{The Luttinger formula (\Eq{sigmaK}) is wrong}
The Comment argues that the derivation of the Luttinger formula from Kubo formalism is incorrect because of the use of the \Comment{nonrelativistic realization of causality has been used reflected in the one-sided Fourier transformation of all relevant quantities}. It further states that the \Comment{quantum field theoretical formalism appropriate in this case employs the Feynman Green functions, which originate from a time-ordered product of field operators}. However, inspection of \cite{PabloMauroComparisonKuboQFT2024}, in particular of Eq. (77), which presents the Fourier transform used, and the text after Eq. (102) shows that the standard (two-sided) Fourier transform of causal transport coefficients and the Matsubara time-ordered Green function (the Feynman Green function in imaginary time) are properly used in \cite{PabloMauroComparisonKuboQFT2024}. Therefore, these issues exposed in \cite{CommentPRB} are simply not present in \cite{PabloMauroComparisonKuboQFT2024}, as an inspection of the text shows, and the derivation of the electric conductivity as the Luttinger formula from the application of Kubo formalism to the Ohm law remains correct.

%\subsection{The use of losses $\Gamma > 0$ for graphene is inherently wrong, and goes beyond the description of graphene inherent to the Dirac model}
In the Comment, it is claimed that \Comment{an incorporation of the dissipation of electronic quasiparticles brings the approach of Ref. \cite{PabloMauroComparisonKuboQFT2024} outside the application region of the Dirac model, where the quasiparticles are considered as noninteracting} and \Comment{By introducing the phenomenological parameter $\Gamma$ for a description the relaxation of quasiparticles in graphene, \cite{PabloMauroComparisonKuboQFT2024} again violated the relativistic description of graphene inherent to the Dirac model}.

In Sect. II.C. of \cite{PabloMauroComparisonKuboQFT2024}, we describe the effect of interactions on the electronic quasiparticles. The existence of losses in graphene (as in all normal materials) cannot be avoided. It is widely accepted, measured and studied in (and beyond) the framework of the approximation of the electronic quasiparticles to the 2D Dirac model, see for example \cite{RevModPhys.81.109, DasSarmaRMP2011,PhysRevLett.99.086804,PhysRevB.76.115434,Gusynin2007b,Gusynin2006b,Gusynin2007,TeberPhDThesis2018,ImpuritiesGraphene}. We consider that the effect of interactions on electronic quasiparticles is described by the self--energy that enters in the Schwinger-Dyson equation for a given microscopic model of interactions \cite{PabloMauroComparisonKuboQFT2024}\cite{ImpuritiesGraphene}. In practice, obtaining this self--energy is rather complicated and in general leads to a frequency--dependent self-energy \cite{ImpuritiesGraphene}; in the simplified model used in \cite{PabloMauroComparisonKuboQFT2024}, this frequency--dependent imaginary part of the self-energy is approached by the (minus) constant phenomenological parameter $\Gamma$ \cite{ImpuritiesGraphene}. $\Gamma >0$ introduces a finite quasiparticle lifetime $\tau = \Gamma^{-1}$, and it enters in the description of the system as an imaginary term in the denominator of the propagator \cite{ImpuritiesGraphene}. This is usually considered a reasonable approximation, widely used by the scientific community \cite{ImpuritiesGraphene}. Note that \cite{ImpuritiesGraphene} discusses about possible issues of the introduction of $\Gamma$ with gauge invariance. In addition, taking into account that the measured electrical conductivity of graphene is a high but finite quantity \cite{NobelPrize:2010-Physics}\cite{Cao2019}, and that the dissipation time has been estimated to be on the order of $\tau\backsim 6\times 10^{-13}\text{ s}$ \cite{DasSarmaRMP2011}\cite{Marinko2009}, we took this non-zero quantity in \cite{PabloMauroComparisonKuboQFT2024}. Note that considering $\Gamma > 0$ instead of $\Gamma=0$ is criticized in the Comment \cite{CommentPRB}, but it is widely used in the literature (see \cite{ImpuritiesGraphene} for example).

%\subsection{\Eq{sigmaK} is not gauge invariant}
In the Comment, it is claimed that \Comment{for the modified expressions for the polarization and conductivity used in \cite{PabloMauroComparisonKuboQFT2024} the transversality condition for $\nu = 0$ is violated. In fact the quantities $\tilde{\Pi}_{\mu\nu}$ and $\sigma^{\rm{K}}_{\mu\nu}$ are not tensors because the subtraction made in Eq. (4) violated a tensorial structure.}
It is important to note here that the covariant generalization of neither $\sigma^{\rm{K}}_{\mu\nu}$ nor $\tilde{\Pi}_{\mu\nu}$ is studied in \cite{PabloMauroComparisonKuboQFT2024}, either explicitly or implicitly. Note that, in the derivation of $\sigma^{\rm{K}}_{\mu\nu}$ using the Kubo formula, the interacting Hamiltonian is proportional to the scalar product of two spatial three-dimensional (3D) vectors: the dipolar moment operator $\hat{d}_{\nu}(\bm{x})$ and the electric field $E^{\nu}(\bm{x},t)$ (Eq. (56) of \cite{PabloMauroComparisonKuboQFT2024}). Consequently, the indices of $\sigma^{\rm{K}}_{\mu\nu}$ only can be of spatial character, and the $0$--components are not defined in \cite{PabloMauroComparisonKuboQFT2024}. As a consequence, the gauge dependence of the covariant generalization of the Luttinger formula proposed in the Comment does not pertain to the work presented in \cite{PabloMauroComparisonKuboQFT2024}. Moreover, in \cite{PabloMauroComparisonKuboQFT2024}, this particular covariant generalization is not considered for $\sigma^{\rm{K}}_{\mu\nu}$, see in Sect. IV, page 11 "$\Pi_{\mu\nu}$ is a tensor that relates two four-vectors, while $\sigma_{\mu\nu}$ relates one four-vector with a part of a second-order tensor".
About the tensor nature of $\sigma^{\rm{K}}_{\mu\nu}$, note that the Ohm law $J_{\mu} = \sigma_{\mu\nu}E^{\nu}$ relates the spatial electric current vector $J_{\mu}$ to the spatial electric vector $E^{\nu}$. Therefore, under a change of spatial variables, $\sigma_{\mu\nu}$ behaves as a spatial 2-rank tensor in 3D space.
Additionally, the covariant generalization of $\sigma^{\rm{K}}_{\mu\nu}$ is not discussed in \cite{PabloMauroComparisonKuboQFT2024}. The particular generalization proposed in the Comment (but not in \cite{PabloMauroComparisonKuboQFT2024}) of $\sigma^{\rm{K}}_{\mu\nu}$ from spatial to space--time indices is not supported by the results shown in \cite{PabloMauroComparisonKuboQFT2024}. In conclusion, the Comment critics an extension of the expression discussed in \cite{PabloMauroComparisonKuboQFT2024}, which is introduced in the Comment in \cite{CommentPRB}, but not considered in \cite{PabloMauroComparisonKuboQFT2024}. The actual expression for $\sigma^{\rm{K}}_{\mu\nu}$ discussed in \cite{PabloMauroComparisonKuboQFT2024} remains valid both mathematically and physically, and the claim of broken the gauge invariance done in \cite{CommentPRB} does not apply to the work presented in \cite{PabloMauroComparisonKuboQFT2024}.

%\subsection{The results of the model $\sigma^{\rm{K}}$ do not coincide with the literature.}
In the Comment, it is claimed that \Comment{the results obtained using this modification are physically unacceptable} (meaning the results obtained by the use of the Luttinger formula). In particular that \Comment{for $\omega\geq cq$ \cite{PabloMauroComparisonKuboQFT2024} arrives at
\begin{eqnarray}
\Imag{\sigma_{T}^{\rm{K}}(\omega,\bm{q})} = - \sigma_{0}\dfrac{v_{F}q}{\omega}
\end{eqnarray}
in contradiction with the previously obtained conclusion that at zero temperature the conductivity of pure graphene in the region of propagating waves is real at all frequencies \cite{Falkovsky2007,KatsnelsonBook}.}
It is worth stressing that both Ref. \cite{Falkovsky2007} and Ref. \cite{KatsnelsonBook}, cited in the Comment, use the same Kubo conductivity definition as \cite{PabloMauroComparisonKuboQFT2024} (see Eq. 8 in \cite{Falkovsky2007} and the preceding paragraph, and Eqs. (7.48) -- (7.52) in \cite{KatsnelsonBook}), which differs from the definition used by the Authors of the Comment.
In \cite{Falkovsky2007} the \textbf{local} and \textbf{zero temperature} limits ($k\to0$ and $T\to0$ limits) are explored to obtain a limiting value for the conductivity in Eq. (12) of \cite{Falkovsky2007} for $kv_{0}\ll T\ll \omega$ (note that $v_{F}$ and $q$ in \cite{PabloMauroComparisonKuboQFT2024} correspond to $v_{0}$ and $k$ in \cite{Falkovsky2007}). If we also use the same local limit for studying the value of $\Imag{\sigma_{T}^{\rm K}(\omega,\bm{q})}$, it is easy to see that it also equals zero in the corresponding limit $\dlim_{q\to0}\dlim_{T\to0}\Imag{\sigma_{T}^{\rm K}(\omega,\bm{q})} = 0$.
Concerning the comparison with the book \cite{KatsnelsonBook}, all results in \cite{KatsnelsonBook} are confirmed limiting cases of expressions derived in \cite{PabloMauroComparisonKuboQFT2024}. In particular, if we look at Eqs. (7.36) and (7.38), they show that $\Imag{\sigma(\omega)}=0$ in the local limit for pristine graphene ($\mu=0$), which is what one obtains by setting $q=0$ in Eq. (15) of the Comment \cite{CommentPRB}.
Finally, in the limit studied in \cite{Falkovsky2007} $kv_{0}\ll T\ll\omega$ ($\hbar v_{F}q\ll k_{B}T\ll\hbar\omega$ with the notation of \cite{PabloMauroComparisonKuboQFT2024}), the result in Eq. (18) of \cite{CommentPRB} can be written using asymptotic notation as 
\begin{eqnarray}
\Imag{\sigma_{T}^{\rm K}(\omega,\bm{q})} = 0 + \mathcal{O}\left(\frac{v_Fq}{\omega},\dfrac{k_{B}T}{\hbar\omega}\right),
%\label{eq24}
\end{eqnarray}
being exactly zero in the corresponding limits $\frac{v_Fq}{\omega},\tfrac{k_{B}T}{\hbar\omega} \ll 1$ when the appropriate mathematical study of the limit is correctly done.
Therefore, there is no contradiction between \cite{PabloMauroComparisonKuboQFT2024} and any other paper (\cite{Falkovsky2007} and \cite{KatsnelsonBook}) cited in \cite{CommentPRB} using the same standard Kubo formula.

%\subsection{The model $\sigma^{\rm{K}}$ (as the model $\sigma^{\rm{NR}}$) predicts $J_{\mu}\neq 0$ for $E^{\nu} = 0$, therefore, the Luttinger formula also fails in this purpose.}
In the Comment, it is claimed that \Comment{at nonzero temperature in the region of propagating waves the pure imaginary current in the absence of electric field arises both in the quantum field theoretical formalism using Eq. (20) and in the approach of \cite{PabloMauroComparisonKuboQFT2024}.}
As evidence, they show that, for propagating waves ($cq\leq \omega$), in the $\omega\to0$ limit, it is obtained in the two formalisms that (Eq. (29) of \cite{CommentPRB})
\begin{eqnarray}\label{plasmapeak}
\dlim_{\omega\to0}\dlim_{cq\leq\omega}\Imag{\sigma_{P}(\omega,\bm{q})} = \sigma_{0}\dfrac{\omega_{P}(\bm{q})}{\omega}
\end{eqnarray}
where $\sigma_{0} = \tfrac{e^{2}}{4\hbar}$ is the universal conductivity of graphene, $P\in\{L,T\}$ and $\hbar\omega_{P}(\bm{q}) = \dfrac{8\log(2)}{\pi}\big(k_{B}T - \hbar v_{F}q\delta_{P,T}\big)$. Note that in the purely local limit, the last term can be disregarded. The Authors of the Comment explicitly state that a purely electrically induced current (described by $J_{\mu} = \sigma_{\mu\nu}E^{\nu}$) can exist in a normal material without any applied electric field. We disagree with this statement, we rigorously demonstrated in \cite{PabloMauroComparisonKuboQFT2024} that this cannot happen for $\sigma^{\rm{K}}_{\mu\nu}$, and that a purely electrically generated current needs a non-zero electric field.

To clarify this point, we distinguish the contribution of each peak in three cases: (a) the Drude peak, present when $\Gamma > 0$; (b) the propagating plasma peak of $\sigma^{\rm{K}}_{\mu\nu}$, which arises in the $\Gamma \to 0$ limit; and (c) the anomalous third peak of $\sigma^{\rm{NR}}_{\mu\nu}$, which originates from a magnetically induced current. We treat each in turn.

(a) Drude peak ($\Gamma > 0$). We confirm that, close to the local limit ($\bm{q}\to\bm{0}$) and for $\mu=0$, the intraband conductivity takes the form
\begin{eqnarray}\label{Drudepeak}
\sigma_{P}^{\rm{K},\text{intra}}(\omega,\bm{q}\to\bm{0})
= \sigma_{0}\dfrac{\omega_{p}(\bm{0})}{-\ii(\omega + \ii\Gamma)}
 \xrightarrow[\Gamma\to 0]{}  \ii\sigma_{0}\dfrac{\omega_{p}(\bm{0})}{\omega}
\end{eqnarray}
and it is transformed into the plasma peak shown in \Eq{plasmapeak} when $\Gamma\to0$. We will see in what follows that, for a Drude peak, the losses $\Gamma > 0$ will cause the electric current to dissipate over time.

(b) Propagating plasma peak ($\Gamma\to0$). In this limit the Drude peak transforms into the plasma peak of \Eq{plasmapeak}. Contrary to what is claimed in the Comment, this plasma peak does not imply that purely electrically generated currents exist in the absence of an electric field. To see this, let us start from \Eq{plasmapeak} and show explicitly what the induced electric current is in this case.

First of all, using the Kramers-Kronig relation of \Eq{plasmapeak}, we obtain
\begin{eqnarray}
\Real{\sigma_{P}(\omega,\bm{q}\to\bm{0})} = \sigma_{0}\pi\omega_{P}(\bm{0})\delta(\omega).
\end{eqnarray}
Therefore, contrary to the claim of the Comment, an imaginary conductivity generates a non-zero real DC electric conductivity with direct experimental relevance. This is no more than the Sokhotski-Plemelj formula
\begin{eqnarray}
\dlim_{\Gamma\to0^{+}}\dfrac{1}{-\ii(\omega + \ii\Gamma)} = \pi\delta(\omega) + \ii\mathcal{P}\Big[\tfrac{1}{\omega}\Big],
\end{eqnarray}
from where the $\Gamma = 0$ limit is studied here. Using that
\begin{eqnarray}
\int_{-\infty}^{\infty}\dfrac{\dd\omega}{2\pi}\dfrac{e^{-\ii\omega t}}{-\ii(\omega + \ii\Gamma)} = e^{-\Gamma t}\Theta(t),
\end{eqnarray}
where $\Theta(t)$ is the Heaviside Theta function (note that the standard 2-sided Fourier transform is always used in \cite{PabloMauroComparisonKuboQFT2024}), the induced electric current in position space is
\begin{eqnarray}
J_{x}(t,\bm{0})
& = & \int_{-\infty}^{\infty}\dd\tau \sigma_{0}\omega_{P}(\bm{0})e^{-\Gamma (t-\tau)}\Theta(t - \tau)E_{x}(\tau,\bm{0})\nonumber\\
& = & \sigma_{0}\omega_{P}(\bm{0})\int_{-\infty}^{t}\dd \tau e^{-\Gamma (t-\tau)}E_{x}(\tau,\bm{0}).
\end{eqnarray}
Note that, for a Drude peak, the losses $\Gamma >0$ make the electric current dissipate with time. However, for the plasma peak, as $\Gamma = 0$, we have a non-dissipative induced electric current $J_{x}(t,\bm{0})$, induced by an electric field $E_{x}(\tau,\bm{0})$. In particular, if the electric field is switched off at a time $t_{0}<t$, for the Drude case, $J_{x}(t)$ will dissipate, while for the plasma case, $J_{x}(t)$ remains constant without dissipating as a permanent current. This is another reason why $\Gamma > 0$ has to be imposed for properly describing the electric transport properties of graphene. In either of those two cases, if $E_{x}(\tau) = 0\,\forall \tau\leq t$, it is obtained that $J_{x}(t) = 0 \,\forall\, t\in\mathbb{R}$ by causality.
In conclusion, and contrary to the claims of the Authors of the Comment, a plasma peak does not imply the existence of an induced electric current in absence of an electric field. Therefore neither the model $\sigma^{\rm{K}}$ nor the two plasma peaks of the propagating sector of $\sigma^{\rm{NR}}$ have this physical inconsistency.

(c) Anomalous third peak of $\sigma^{\rm{NR}}_{\mu\nu}$. However, in \cite{PabloMauroComparisonKuboQFT2024} it is claimed that \Eq{sigmaNR} generates an electric current in the absence of an electric field, and the authors of the Comment agree on this point, when they claimed \Comment{at nonzero temperature in both the quantum field theory and in the formalism of \cite{PabloMauroComparisonKuboQFT2024} a nonzero current in graphene arises even for a zero electric field}, is it not contradictory with this behavior of a plasma peak for propagating waves pointed out by the authors of the Comment?

Not really. This apparent tension is resolved once one recognizes that the propagating plasma peaks and the anomalous third peak in the electric conductivity described by \Eq{sigmaNR} have different physical origins. To see why, note that \Eq{sigmaNR} is derived using the relation (valid in temporal gauge) $E^{\nu}(\omega,\bm{q}) = \ii\omega A^{\nu}(\omega,\bm{q})$ to the assumed true equality
\begin{eqnarray}\label{Relation_Pi_sigmaNR}
J_{\mu}(\omega,\bm{q})
& = & \Pi_{\mu\nu}(\omega,\bm{q})A^{\nu}(\omega,\bm{q}) = \dfrac{\Pi_{\mu\nu}(\omega,\bm{q})}{\ii\omega}\Big[\ii\omega A^{\nu}(\omega,\bm{q})\Big]\nonumber\\
& = & \sigma_{\mu\nu}^{\rm{NR}}(\omega,\bm{q})E^{\nu}(\omega,\bm{q}).
\end{eqnarray}
The problem is that, contrary to the assumption of the Comment of being \Comment{in the absence of the constant in time, external magnetic field}, in \cite{PabloMauroComparisonKuboQFT2024}, like in \Eq{Relation_Pi_sigmaNR}, we can always choose $A^{\nu}(\bm{x},t) = \delta^{\nu 2}B_{0}x$ for a constant (in time) magnetic field $B^{\nu} = B_{0}\delta^{\nu 3}$ that induces an electric current by using
\begin{eqnarray}\label{def_Pi_w=0}
J_{\mu}(0,\bm{q}) = \Pi_{\mu\nu}(0,\bm{q})A^{\nu}(0,\bm{q}).
\end{eqnarray}
However, it is important to note that, for $A^{\nu}(\bm{x},t) = \delta^{\nu 2}B_{0}x$, one has $E^{\nu}(\omega,\bm{q}) = \ii\omega A^{\nu}(\omega,\bm{q}) = 0$. Moreover, a constant (in time) magnetic field cannot be expressed in terms of an electric field by using the Faraday law, because $\partial_{t}\bm{B} = - \bm{\nabla}\times\bm{E} = \bm{0}$. Therefore, in this particular case, neither $A^{\nu}(\omega,\bm{q}) = E^{\nu}(\omega,\bm{q})/(\ii\omega)$ nor the relation
\begin{eqnarray}\label{FaradayLaw}
\bm{B}(\omega,\bm{q}) = \dfrac{1}{\omega}\bm{q}\times\bm{E}(\omega,\bm{q})
\end{eqnarray}
can be used to write the vector potential or magnetic field in terms of the electric field in \Eq{Relation_Pi_sigmaNR}. If we ignore this fact, and incorrectly applies the Faraday Law (\Eq{FaradayLaw}) to this constant magnetic field, the induced electric current is described by
\begin{eqnarray}
\sigma^{\rm{NR},\text{inter}}_{T}(\omega,\bm{q}) = \dfrac{\dlim_{\omega\to0}\Pi_{T}(\omega,\bm{q})}{\ii\omega},
\end{eqnarray}
which is exactly the problematic plasma peak that appears in the model of the Authors of the Comment of $\sigma^{\rm{NR}}$ \cite{PabloMauroComparisonKuboQFT2024}. Therefore, we have an induced electric current described by the Ohm Law, but in absence of an electric field (induced by a constant magnetic field). This is the third anomalous plasma peak of $\sigma^{\rm{NR}}$, a peak in the evanescent sector, coming from the interband term of the transversal conductivity that cannot be eliminated by the presence of losses $\Gamma > 0$ and that really describes the induced electric current by a constant magnetic field.
Note that the model for $\sigma^{\rm{K}}$ shown in \cite{PabloMauroComparisonKuboQFT2024} naturally avoids this problem without any ad-hoc modification, as a mathematical derived consequence. By contrast, the model of the authors of the Comment does not, as they admit in their Comment. See Sect. IIIB of \cite{PabloMauroComparisonKuboQFT2024}, where it is shown that, if the electric field is zero, the induced electric current defined as $J_{\mu} = \sigma^{\rm{K}}_{\mu\nu}E^{\nu}$ is zero as well.

%\subsection{Results of the comment in absence of constant magnetic field.}
An important limitation of the Comment's argument must be stressed: the Authors restrict their analysis to the case \Comment{in the absence of constant in time, external magnetic field}, a restriction that is not imposed in \cite{PabloMauroComparisonKuboQFT2024}, where any small electromagnetic perturbation is considered, including constant magnetic fields.

Therefore, any statement in the Comment that relies on the absence of static magnetic field is not applicable to \cite{PabloMauroComparisonKuboQFT2024}.
In particular, the claim that \Eq{Def_Pi} (Eq. (20) of \cite{CommentPRB} $J_{\mu} = - \Pi_{\mu\nu}A^{\nu}$) is considered unsatisfactory is incorrect. In \cite{PabloMauroComparisonKuboQFT2024}, \Eq{Def_Pi} is treated as the most general relation between induced electric currents and electromagnetic fields (including any arbitrary magnetic field that can be described as a linear perturbation of the system). \Eq{Def_Pi} is studied also with the Kubo formalism in \cite{PabloMauroComparisonKuboQFT2024}, and known results (of the Authors of the Comment and others) are recovered. The same applies to Eq. (31) of the Comment (\Eq{sigmaNR} here) which, contrary to what is claimed in the Comment, contains magnetically induced parts. The point is that those contributions have not been subtracted from \Eq{sigmaNR}.

%\subsection{There are several typos and the system of units is not clear.}
In the Comment, it is claimed that the system of units is not clear. We clarify that SI units are used throughout, with no fundamental constant set equal to unity.

Finally, we take the opportunity in this Reply to be correct few minor typos in \cite{PabloMauroComparisonKuboQFT2024}: $\alpha c$ should be replaced by $e^{2}/\hbar$, as we already pointed out in \cite{PabloMauro2025}; in Table I, $\kk$ should be defined using $\tilde{k}_{0}$ in the place of $k_{0}$; instead of ``the case $\Pi_{\mu\nu}(\qq) = \Pi_{L}(\qq)$'', it should read ``For the longitudinal part of $\Pi_{\mu\nu}(\qq)$, denoted as $\Pi_{L}(\qq)$''; Throughout \cite{PabloMauroComparisonKuboQFT2024}, $q_{z}$ is defined as $q_{z} = \sqrt{(\frac{\omega}{v_{F}})^{2} - q_{\surf}^{2}}$. However, in Eqs. (120)-(122), a locally defined quantity $q_{z} = \sqrt{(\frac{\omega}{c})^{2}  - q_{\surf}^{2}}$ is used, as stated explicitly in the surrounding text.

%\subsection{epsilon}
In the Comment it is claimed that \Comment{As to a prediction of the double pole at zero frequency by quantum field theory, it does not contradict to any physical results, including the Kubo approach formulated for the propagating fields}. However, this observation misses the physical point raised in \cite{PabloMauroComparisonKuboQFT2024}: the anomalous plasma peak responsible for the double pole comes from a magnetic contribution to the induced electric current. Accordingly, this term contributes to a magnetic rather than dielectric permittivity of the material.

%\subsection{Thermal Casimir}
In the Comment, it is claimed that \Comment{Note also that just the behavior of the conductivities of graphene according to \Eq{plasmapeak} (Eq. (26) of the Comment) leads to the big thermal effect in the Casimir force between two graphene sheets at short separations predicted in \cite{GomezSantos2009} for the case of two pristine graphene sheets and confirmed experimentally in \cite{PRL_Mohideen,PRB_Mohideen}.}, implying independent experimental support for their model. Moreover, even aside from not constituting a critique of \cite{PabloMauroComparisonKuboQFT2024}, this statement is incorrect. Actually, \Eq{plasmapeak} is not needed to obtain \Comment{the big thermal effect in the Casimir force between two graphene sheets at short separations predicted in \cite{GomezSantos2009}}. This result is obtained from any 2D material with $\dlim_{\omega\to0}\sigma_{L}(\omega)>0$ (which holds whenever $\Gamma > 0$), as we recently showed in \cite{PabloMauro2025}. Therefore, this big thermal Casimir effect for graphene is not only derived from \Eq{plasmapeak}, but it is also a general property of any 2D Drude metal, and the experiments shown in \cite{PRL_Mohideen,PRB_Mohideen} cannot serve as a discriminating test of \Eq{plasmapeak} of the Comment.

%\section{Conclusions}
We have replied to all concerns raised in the Comment \cite{CommentPRB}, we show that they do not affect our conclusions, and that all discussions, derivations and results of \cite{PabloMauroComparisonKuboQFT2024} remain fully valid.

\begin{acknowledgments}
P. R.-L. acknowledges support from Ministerio de Ciencia, Innovaci\'on y Universidades (Spain), Agencia Estatal de Investigaci\'on, under project NAUTILUS (PID2022-139524NB-I00), from AYUDA PUENTE, URJC, from QuantUM program of the University of Montpellier and the hospitality of the Theory of Light-Matter and Quantum Phenomena group at the Laboratoire Charles Coulomb, University of Montpellier, where part of this work was done.
J.-S.W. acknowledges support from MOE FRC tier 1 grant A-8000990-00-00.
M.A. acknowledges the QuantUM program of the University of Montpellier, the grant ”CAT”, No. A-HKUST604/20, from the ANR/RGC Joint Research Scheme sponsored by the French National Research Agency (ANR) and the Research Grants Council (RGC) of the Hong Kong Special Administrative Region.
%P.R.-L., and M.A. acknowledge the QuantUM program of the University of Montpellier
\end{acknowledgments}

\newpage
\providecommand{\noopsort}[1]{}\providecommand{\singleletter}[1]{#1}%


\begin{thebibliography}{21}%
\makeatletter
\providecommand \@ifxundefined [1]{%
 \@ifx{#1\undefined}
}%
\providecommand \@ifnum [1]{%
 \ifnum #1\expandafter \@firstoftwo
 \else \expandafter \@secondoftwo
 \fi
}%
\providecommand \@ifx [1]{%
 \ifx #1\expandafter \@firstoftwo
 \else \expandafter \@secondoftwo
 \fi
}%
\providecommand \natexlab [1]{#1}%
\providecommand \enquote  [1]{``#1''}%
\providecommand \bibnamefont  [1]{#1}%
\providecommand \bibfnamefont [1]{#1}%
\providecommand \citenamefont [1]{#1}%
\providecommand \href@noop [0]{\@secondoftwo}%
\providecommand \href [0]{\begingroup \@sanitize@url \@href}%
\providecommand \@href[1]{\@@startlink{#1}\@@href}%
\providecommand \@@href[1]{\endgroup#1\@@endlink}%
\providecommand \@sanitize@url [0]{\catcode `\\12\catcode `\$12\catcode
  `\&12\catcode `\#12\catcode `\^12\catcode `\_12\catcode `\%12\relax}%
\providecommand \@@startlink[1]{}%
\providecommand \@@endlink[0]{}%
\providecommand \url  [0]{\begingroup\@sanitize@url \@url }%
\providecommand \@url [1]{\endgroup\@href {#1}{\urlprefix }}%
\providecommand \urlprefix  [0]{URL }%
\providecommand \Eprint [0]{\href }%
\providecommand \doibase [0]{https://doi.org/}%
\providecommand \selectlanguage [0]{\@gobble}%
\providecommand \bibinfo  [0]{\@secondoftwo}%
\providecommand \bibfield  [0]{\@secondoftwo}%
\providecommand \translation [1]{[#1]}%
\providecommand \BibitemOpen [0]{}%
\providecommand \bibitemStop [0]{}%
\providecommand \bibitemNoStop [0]{.\EOS\space}%
\providecommand \EOS [0]{\spacefactor3000\relax}%
\providecommand \BibitemShut  [1]{\csname bibitem#1\endcsname}%
\let\auto@bib@innerbib\@empty
%</preamble>
\bibitem [{\citenamefont {Bordag}\ \emph {et~al.}(2026)\citenamefont {Bordag},
  \citenamefont {Khusnutdinov}, \citenamefont {Klimchitskaya},\ and\
  \citenamefont {Mostepanenko}}]{CommentPRB}%
  \BibitemOpen
  \bibfield  {author} {\bibinfo {author} {\bibfnamefont {M.}~\bibnamefont
  {Bordag}}, \bibinfo {author} {\bibfnamefont {N.}~\bibnamefont
  {Khusnutdinov}}, \bibinfo {author} {\bibfnamefont {G.~L.}\ \bibnamefont
  {Klimchitskaya}},\ and\ \bibinfo {author} {\bibfnamefont {V.~M.}\
  \bibnamefont {Mostepanenko}},\ }\bibfield  {title} {\bibinfo {title} {Comment
  on ``electric conductivity in graphene: Kubo model versus a nonlocal quantum
  field theory model''},\ }\href {https://doi.org/10.1103/pvgr-rf1z} {\bibfield
   {journal} {\bibinfo  {journal} {Phys. Rev. B}\ }\textbf {\bibinfo {volume}
  {113}},\ \bibinfo {pages} {207401} (\bibinfo {year} {2026})}\BibitemShut
  {NoStop}%
\bibitem [{\citenamefont {Rodriguez-Lopez}\ \emph {et~al.}(2025)\citenamefont
  {Rodriguez-Lopez}, \citenamefont {Wang},\ and\ \citenamefont
  {Antezza}}]{PabloMauroComparisonKuboQFT2024}%
  \BibitemOpen
  \bibfield  {author} {\bibinfo {author} {\bibfnamefont {P.}~\bibnamefont
  {Rodriguez-Lopez}}, \bibinfo {author} {\bibfnamefont {J.-S.}\ \bibnamefont
  {Wang}},\ and\ \bibinfo {author} {\bibfnamefont {M.}~\bibnamefont
  {Antezza}},\ }\bibfield  {title} {\bibinfo {title} {Electric conductivity in
  graphene: Kubo model versus a nonlocal quantum field theory model},\ }\href
  {https://doi.org/10.1103/PhysRevB.111.115428} {\bibfield  {journal} {\bibinfo
   {journal} {Phys. Rev. B}\ }\textbf {\bibinfo {volume} {111}},\ \bibinfo
  {pages} {115428} (\bibinfo {year} {2025})}\BibitemShut {NoStop}%
\bibitem [{\citenamefont {Rodriguez-Lopez}\ and\ \citenamefont
  {Antezza}(2025)}]{PabloMauro2025}%
  \BibitemOpen
  \bibfield  {author} {\bibinfo {author} {\bibfnamefont {P.}~\bibnamefont
  {Rodriguez-Lopez}}\ and\ \bibinfo {author} {\bibfnamefont {M.}~\bibnamefont
  {Antezza}},\ }\bibfield  {title} {\bibinfo {title} {Casimir-lifshitz force
  with graphene: Role of spatial nonlocality and of losses},\ }\href
  {https://doi.org/10.1103/vwm1-x3vw} {\bibfield  {journal} {\bibinfo
  {journal} {Phys. Rev. B}\ }\textbf {\bibinfo {volume} {112}},\ \bibinfo
  {pages} {035412} (\bibinfo {year} {2025})}\BibitemShut {NoStop}%
\bibitem [{\citenamefont {Castro~Neto}\ \emph {et~al.}(2009)\citenamefont
  {Castro~Neto}, \citenamefont {Guinea}, \citenamefont {Peres}, \citenamefont
  {Novoselov},\ and\ \citenamefont {Geim}}]{RevModPhys.81.109}%
  \BibitemOpen
  \bibfield  {author} {\bibinfo {author} {\bibfnamefont {A.~H.}\ \bibnamefont
  {Castro~Neto}}, \bibinfo {author} {\bibfnamefont {F.}~\bibnamefont {Guinea}},
  \bibinfo {author} {\bibfnamefont {N.~M.~R.}\ \bibnamefont {Peres}}, \bibinfo
  {author} {\bibfnamefont {K.~S.}\ \bibnamefont {Novoselov}},\ and\ \bibinfo
  {author} {\bibfnamefont {A.~K.}\ \bibnamefont {Geim}},\ }\bibfield  {title}
  {\bibinfo {title} {The electronic properties of graphene},\ }\href
  {https://doi.org/10.1103/RevModPhys.81.109} {\bibfield  {journal} {\bibinfo
  {journal} {Rev. Mod. Phys.}\ }\textbf {\bibinfo {volume} {81}},\ \bibinfo
  {pages} {109} (\bibinfo {year} {2009})}\BibitemShut {NoStop}%
\bibitem [{\citenamefont {Das~Sarma}\ \emph {et~al.}(2011)\citenamefont
  {Das~Sarma}, \citenamefont {Adam}, \citenamefont {Hwang},\ and\ \citenamefont
  {Rossi}}]{DasSarmaRMP2011}%
  \BibitemOpen
  \bibfield  {author} {\bibinfo {author} {\bibfnamefont {S.}~\bibnamefont
  {Das~Sarma}}, \bibinfo {author} {\bibfnamefont {S.}~\bibnamefont {Adam}},
  \bibinfo {author} {\bibfnamefont {E.~H.}\ \bibnamefont {Hwang}},\ and\
  \bibinfo {author} {\bibfnamefont {E.}~\bibnamefont {Rossi}},\ }\bibfield
  {title} {\bibinfo {title} {Electronic transport in two-dimensional
  graphene},\ }\href {https://doi.org/10.1103/RevModPhys.83.407} {\bibfield
  {journal} {\bibinfo  {journal} {Rev. Mod. Phys.}\ }\textbf {\bibinfo {volume}
  {83}},\ \bibinfo {pages} {407} (\bibinfo {year} {2011})}\BibitemShut
  {NoStop}%
\bibitem [{\citenamefont {Park}\ \emph {et~al.}(2007)\citenamefont {Park},
  \citenamefont {Giustino}, \citenamefont {Cohen},\ and\ \citenamefont
  {Louie}}]{PhysRevLett.99.086804}%
  \BibitemOpen
  \bibfield  {author} {\bibinfo {author} {\bibfnamefont {C.-H.}\ \bibnamefont
  {Park}}, \bibinfo {author} {\bibfnamefont {F.}~\bibnamefont {Giustino}},
  \bibinfo {author} {\bibfnamefont {M.~L.}\ \bibnamefont {Cohen}},\ and\
  \bibinfo {author} {\bibfnamefont {S.~G.}\ \bibnamefont {Louie}},\ }\bibfield
  {title} {\bibinfo {title} {Velocity renormalization and carrier lifetime in
  graphene from the electron-phonon interaction},\ }\href
  {https://doi.org/10.1103/PhysRevLett.99.086804} {\bibfield  {journal}
  {\bibinfo  {journal} {Phys. Rev. Lett.}\ }\textbf {\bibinfo {volume} {99}},\
  \bibinfo {pages} {086804} (\bibinfo {year} {2007})}\BibitemShut {NoStop}%
\bibitem [{\citenamefont {Hwang}\ \emph {et~al.}(2007)\citenamefont {Hwang},
  \citenamefont {Hu},\ and\ \citenamefont {Das~Sarma}}]{PhysRevB.76.115434}%
  \BibitemOpen
  \bibfield  {author} {\bibinfo {author} {\bibfnamefont {E.~H.}\ \bibnamefont
  {Hwang}}, \bibinfo {author} {\bibfnamefont {B.~Y.-K.}\ \bibnamefont {Hu}},\
  and\ \bibinfo {author} {\bibfnamefont {S.}~\bibnamefont {Das~Sarma}},\
  }\bibfield  {title} {\bibinfo {title} {Inelastic carrier lifetime in
  graphene},\ }\href {https://doi.org/10.1103/PhysRevB.76.115434} {\bibfield
  {journal} {\bibinfo  {journal} {Phys. Rev. B}\ }\textbf {\bibinfo {volume}
  {76}},\ \bibinfo {pages} {115434} (\bibinfo {year} {2007})}\BibitemShut
  {NoStop}%
\bibitem [{\citenamefont {Gusynin}\ \emph {et~al.}(2006)\citenamefont
  {Gusynin}, \citenamefont {Sharapov},\ and\ \citenamefont
  {Carbotte}}]{Gusynin2007b}%
  \BibitemOpen
  \bibfield  {author} {\bibinfo {author} {\bibfnamefont {V.~P.}\ \bibnamefont
  {Gusynin}}, \bibinfo {author} {\bibfnamefont {S.~G.}\ \bibnamefont
  {Sharapov}},\ and\ \bibinfo {author} {\bibfnamefont {J.~P.}\ \bibnamefont
  {Carbotte}},\ }\bibfield  {title} {\bibinfo {title} {Magneto-optical
  conductivity in graphene},\ }\href
  {https://doi.org/10.1088/0953-8984/19/2/026222} {\bibfield  {journal}
  {\bibinfo  {journal} {Journal of Physics: Condensed Matter}\ }\textbf
  {\bibinfo {volume} {19}},\ \bibinfo {pages} {026222} (\bibinfo {year}
  {2006})}\BibitemShut {NoStop}%
\bibitem [{\citenamefont {Gusynin}\ and\ \citenamefont
  {Sharapov}(2006)}]{Gusynin2006b}%
  \BibitemOpen
  \bibfield  {author} {\bibinfo {author} {\bibfnamefont {V.~P.}\ \bibnamefont
  {Gusynin}}\ and\ \bibinfo {author} {\bibfnamefont {S.~G.}\ \bibnamefont
  {Sharapov}},\ }\bibfield  {title} {\bibinfo {title} {Transport of dirac
  quasiparticles in graphene: Hall and optical conductivities},\ }\href
  {https://doi.org/10.1103/PhysRevB.73.245411} {\bibfield  {journal} {\bibinfo
  {journal} {Phys. Rev. B}\ }\textbf {\bibinfo {volume} {73}},\ \bibinfo
  {pages} {245411} (\bibinfo {year} {2006})}\BibitemShut {NoStop}%
\bibitem [{\citenamefont {Gusynin}\ \emph {et~al.}(2007)\citenamefont
  {Gusynin}, \citenamefont {Sharapov},\ and\ \citenamefont
  {Cabotte}}]{Gusynin2007}%
  \BibitemOpen
  \bibfield  {author} {\bibinfo {author} {\bibfnamefont {V.~P.}\ \bibnamefont
  {Gusynin}}, \bibinfo {author} {\bibfnamefont {S.~G.}\ \bibnamefont
  {Sharapov}},\ and\ \bibinfo {author} {\bibfnamefont {J.~P.}\ \bibnamefont
  {Cabotte}},\ }\bibfield  {title} {\bibinfo {title} {Ac conductivity of
  graphene: From tight-binding model to 2 + 1-dimensional quantum
  electrodynamics},\ }\href {https://doi.org/10.1142/S0217979207038022}
  {\bibfield  {journal} {\bibinfo  {journal} {International Journal of Modern
  Physics B}\ }\textbf {\bibinfo {volume} {21}},\ \bibinfo {pages} {4611}
  (\bibinfo {year} {2007})},\ \Eprint
  {https://arxiv.org/abs/https://doi.org/10.1142/S0217979207038022}
  {https://doi.org/10.1142/S0217979207038022} \BibitemShut {NoStop}%
\bibitem [{\citenamefont {Teber}(2018)}]{TeberPhDThesis2018}%
  \BibitemOpen
  \bibfield  {author} {\bibinfo {author} {\bibfnamefont {S.}~\bibnamefont
  {Teber}},\ }\emph {\bibinfo {title} {Field theoretic study of
  electron-electron interaction effects in Dirac liquids}},\ \href
  {https://arxiv.org/abs/1810.08428} {\bibinfo {type} {Phd thesis}},\ \bibinfo
  {school} {Sorbonne Universite, Paris, France} (\bibinfo {year} {2018}),\
  \Eprint {https://arxiv.org/abs/1810.08428} {arXiv:1810.08428
  [cond-mat.mes-hall]} \BibitemShut {NoStop}%
\bibitem [{\citenamefont {Khusnutdinov}\ and\ \citenamefont
  {Vassilevich}(2024)}]{ImpuritiesGraphene}%
  \BibitemOpen
  \bibfield  {author} {\bibinfo {author} {\bibfnamefont {N.}~\bibnamefont
  {Khusnutdinov}}\ and\ \bibinfo {author} {\bibfnamefont {D.}~\bibnamefont
  {Vassilevich}},\ }\bibfield  {title} {\bibinfo {title} {Impurities in
  graphene and their influence on the casimir interaction},\ }\href
  {https://doi.org/10.1103/PhysRevB.109.235420} {\bibfield  {journal} {\bibinfo
   {journal} {Phys. Rev. B}\ }\textbf {\bibinfo {volume} {109}},\ \bibinfo
  {pages} {235420} (\bibinfo {year} {2024})}\BibitemShut {NoStop}%
\bibitem [{\citenamefont {Foundation}(2010)}]{NobelPrize:2010-Physics}%
  \BibitemOpen
  \bibfield  {author} {\bibinfo {author} {\bibfnamefont {N.}~\bibnamefont
  {Foundation}},\ }\href
  {https://www.nobelprize.org/prizes/physics/2010/advanced-information/}
  {\bibinfo {title} {Scientific background on the nobel prize in physics 2010}}
  (\bibinfo {year} {2010})\BibitemShut {NoStop}%
\bibitem [{\citenamefont {Cao}\ \emph {et~al.}(2019)\citenamefont {Cao},
  \citenamefont {Luo}, \citenamefont {Xie}, \citenamefont {Tan}, \citenamefont
  {Fan}, \citenamefont {Guo}, \citenamefont {Su}, \citenamefont {Li},\ and\
  \citenamefont {Xiong}}]{Cao2019}%
  \BibitemOpen
  \bibfield  {author} {\bibinfo {author} {\bibfnamefont {M.}~\bibnamefont
  {Cao}}, \bibinfo {author} {\bibfnamefont {Y.}~\bibnamefont {Luo}}, \bibinfo
  {author} {\bibfnamefont {Y.}~\bibnamefont {Xie}}, \bibinfo {author}
  {\bibfnamefont {Z.}~\bibnamefont {Tan}}, \bibinfo {author} {\bibfnamefont
  {G.}~\bibnamefont {Fan}}, \bibinfo {author} {\bibfnamefont {Q.}~\bibnamefont
  {Guo}}, \bibinfo {author} {\bibfnamefont {Y.}~\bibnamefont {Su}}, \bibinfo
  {author} {\bibfnamefont {Z.}~\bibnamefont {Li}},\ and\ \bibinfo {author}
  {\bibfnamefont {D.-B.}\ \bibnamefont {Xiong}},\ }\bibfield  {title} {\bibinfo
  {title} {The influence of interface structure on the electrical conductivity
  of graphene embedded in aluminum matrix},\ }\href
  {https://doi.org/https://doi.org/10.1002/admi.201900468} {\bibfield
  {journal} {\bibinfo  {journal} {Advanced Materials Interfaces}\ }\textbf
  {\bibinfo {volume} {6}},\ \bibinfo {pages} {1900468} (\bibinfo {year}
  {2019})},\ \Eprint
  {https://arxiv.org/abs/https://onlinelibrary.wiley.com/doi/pdf/10.1002/admi.201900468}
  {https://onlinelibrary.wiley.com/doi/pdf/10.1002/admi.201900468} \BibitemShut
  {NoStop}%
\bibitem [{\citenamefont {Jablan}\ \emph {et~al.}(2009)\citenamefont {Jablan},
  \citenamefont {Buljan},\ and\ \citenamefont {Solja\ifmmode \check{c}\else
  \v{c}\fi{}i\ifmmode~\acute{c}\else \'{c}\fi{}}}]{Marinko2009}%
  \BibitemOpen
  \bibfield  {author} {\bibinfo {author} {\bibfnamefont {M.}~\bibnamefont
  {Jablan}}, \bibinfo {author} {\bibfnamefont {H.}~\bibnamefont {Buljan}},\
  and\ \bibinfo {author} {\bibfnamefont {M.}~\bibnamefont {Solja\ifmmode
  \check{c}\else \v{c}\fi{}i\ifmmode~\acute{c}\else \'{c}\fi{}}},\ }\bibfield
  {title} {\bibinfo {title} {Plasmonics in graphene at infrared frequencies},\
  }\href {https://doi.org/10.1103/PhysRevB.80.245435} {\bibfield  {journal}
  {\bibinfo  {journal} {Phys. Rev. B}\ }\textbf {\bibinfo {volume} {80}},\
  \bibinfo {pages} {245435} (\bibinfo {year} {2009})}\BibitemShut {NoStop}%
\bibitem [{\citenamefont {Falkovsky}\ and\ \citenamefont
  {Varlamov}(2007)}]{Falkovsky2007}%
  \BibitemOpen
  \bibfield  {author} {\bibinfo {author} {\bibfnamefont {L.~A.}\ \bibnamefont
  {Falkovsky}}\ and\ \bibinfo {author} {\bibfnamefont {A.~A.}\ \bibnamefont
  {Varlamov}},\ }\bibfield  {title} {\bibinfo {title} {Space-time dispersion of
  graphene conductivity},\ }\href {https://doi.org/10.1140/epjb/e2007-00142-3}
  {\bibfield  {journal} {\bibinfo  {journal} {The European Physical Journal B}\
  }\textbf {\bibinfo {volume} {56}},\ \bibinfo {pages} {281} (\bibinfo {year}
  {2007})}\BibitemShut {NoStop}%
\bibitem [{\citenamefont {Katsnelson}(2020)}]{KatsnelsonBook}%
  \BibitemOpen
  \bibfield  {author} {\bibinfo {author} {\bibfnamefont {M.~I.}\ \bibnamefont
  {Katsnelson}},\ }\href {https://doi.org/10.1017/9781108617567} {\emph
  {\bibinfo {title} {The Physics of Graphene}}},\ \bibinfo {edition} {2nd}\
  ed.\ (\bibinfo  {publisher} {Cambridge University Press},\ \bibinfo {year}
  {2020})\BibitemShut {NoStop}%
\bibitem [{\citenamefont {G\'omez-Santos}(2009)}]{GomezSantos2009}%
  \BibitemOpen
  \bibfield  {author} {\bibinfo {author} {\bibfnamefont {G.}~\bibnamefont
  {G\'omez-Santos}},\ }\bibfield  {title} {\bibinfo {title} {Thermal van der
  waals interaction between graphene layers},\ }\href
  {https://doi.org/10.1103/PhysRevB.80.245424} {\bibfield  {journal} {\bibinfo
  {journal} {Phys. Rev. B}\ }\textbf {\bibinfo {volume} {80}},\ \bibinfo
  {pages} {245424} (\bibinfo {year} {2009})}\BibitemShut {NoStop}%
\bibitem [{\citenamefont {Liu}\ \emph {et~al.}(2021{\natexlab{a}})\citenamefont
  {Liu}, \citenamefont {Zhang}, \citenamefont {Klimchitskaya}, \citenamefont
  {Mostepanenko},\ and\ \citenamefont {Mohideen}}]{PRL_Mohideen}%
  \BibitemOpen
  \bibfield  {author} {\bibinfo {author} {\bibfnamefont {M.}~\bibnamefont
  {Liu}}, \bibinfo {author} {\bibfnamefont {Y.}~\bibnamefont {Zhang}}, \bibinfo
  {author} {\bibfnamefont {G.~L.}\ \bibnamefont {Klimchitskaya}}, \bibinfo
  {author} {\bibfnamefont {V.~M.}\ \bibnamefont {Mostepanenko}},\ and\ \bibinfo
  {author} {\bibfnamefont {U.}~\bibnamefont {Mohideen}},\ }\bibfield  {title}
  {\bibinfo {title} {Demonstration of an unusual thermal effect in the casimir
  force from graphene},\ }\href
  {https://doi.org/10.1103/PhysRevLett.126.206802} {\bibfield  {journal}
  {\bibinfo  {journal} {Phys. Rev. Lett.}\ }\textbf {\bibinfo {volume} {126}},\
  \bibinfo {pages} {206802} (\bibinfo {year} {2021}{\natexlab{a}})}\BibitemShut
  {NoStop}%
\bibitem [{\citenamefont {Liu}\ \emph {et~al.}(2021{\natexlab{b}})\citenamefont
  {Liu}, \citenamefont {Zhang}, \citenamefont {Klimchitskaya}, \citenamefont
  {Mostepanenko},\ and\ \citenamefont {Mohideen}}]{PRB_Mohideen}%
  \BibitemOpen
  \bibfield  {author} {\bibinfo {author} {\bibfnamefont {M.}~\bibnamefont
  {Liu}}, \bibinfo {author} {\bibfnamefont {Y.}~\bibnamefont {Zhang}}, \bibinfo
  {author} {\bibfnamefont {G.~L.}\ \bibnamefont {Klimchitskaya}}, \bibinfo
  {author} {\bibfnamefont {V.~M.}\ \bibnamefont {Mostepanenko}},\ and\ \bibinfo
  {author} {\bibfnamefont {U.}~\bibnamefont {Mohideen}},\ }\bibfield  {title}
  {\bibinfo {title} {Experimental and theoretical investigation of the thermal
  effect in the casimir interaction from graphene},\ }\href
  {https://doi.org/10.1103/PhysRevB.104.085436} {\bibfield  {journal} {\bibinfo
   {journal} {Phys. Rev. B}\ }\textbf {\bibinfo {volume} {104}},\ \bibinfo
  {pages} {085436} (\bibinfo {year} {2021}{\natexlab{b}})}\BibitemShut
  {NoStop}%
\end{thebibliography}
\end{document}